\iftrue
\documentclass[aps,prl,twocolumn,
			   groupedaddress,superscriptaddress,
			   amsfonts,amssymb,amsmath,
			   citeautoscript,
			   a4paper]{revtex4-1}
\else
\documentclass[aps,pra,reprint,
			   groupedaddress,superscriptaddress,notitlepage,
			   amsfonts,amssymb,amsmath,
			   citeautoscript,
			   a4paper]{revtex4-1}
\fi

\usepackage[pdftex]{hyperref}
\hypersetup{colorlinks,
			linkcolor={blue!75!black!80!yellow},
			citecolor={blue!75!black!80!yellow},
			urlcolor={blue!75!black!80!yellow},
			pdfstartview=FitH}
\usepackage{graphicx}
\usepackage{mathrsfs}
\usepackage{amsthm}
\usepackage{physics}
\usepackage{xspace}
\usepackage{braket}
\usepackage{xr}
\usepackage{gensymb} 
\usepackage{xcolor,soul}
\usepackage{stmaryrd} 
\usepackage[UKenglish]{babel}
\usepackage{placeins} 
\usepackage{siunitx}
\DeclareSIUnit[number-unit-product=]\percent{\char`\%} 
\usepackage{makecell}
\usepackage{diagbox}
\usepackage{float}

\usepackage{txfonts}  
\usepackage{txfontsb} 

\renewcommand{\Re}{\operatorname{Re}}

\newcommand{\iu}{\mathrm{i}}

\newcommand{\appropto}{\mathrel{\vcenter{
			\offinterlineskip\halign{\hfil$##$\cr
				\propto\cr\noalign{\kern2pt}\sim\cr\noalign{\kern-2pt}}}}}

\newcommand{\sigmax}{\sigma_x}
\newcommand{\sigmay}{\sigma_y}
\newcommand{\sigmaz}{\sigma_z}

\newcommand{\ie}{i.e.\@\xspace}  
\newcommand{\cf}{cf.\@\xspace}
\newcommand{\eg}{e.g.\@\xspace}

\makeatletter
\newcommand*{\addFileDependency}[1]{
  \typeout{(#1)}
  \@addtofilelist{#1}
  \IfFileExists{#1}{}{\typeout{No file #1.}}
}
\makeatother

\newcommand*{\myexternaldocument}[1]{%
    \externaldocument{#1}%
    \addFileDependency{#1.tex}%
    \addFileDependency{#1.aux}%
}

\usepackage{scalerel}

\usepackage{textcomp} 
\usepackage{xifthen}
\usepackage{etoolbox}
\newboolean{togglecomments}
\newboolean{togglechanges} 

\setboolean{togglecomments}{true}
\setboolean{togglechanges}{false}

\newcommand{\comment}[2]{%
    \ifbool{togglecomments}%
    {\textcolor{blue!70!black}{\small\textsf{%
    \textsuperscript{\textsc{\textsf{\MakeLowercase{#1}}}}%
    [#2]}}} 
    {}}     
\newcommand{\swap}[2]{\ifbool{togglechanges}
    {#2}  
    {\textcolor{red!70!black}{[#1]}\textrightarrow{}\textcolor{green!50!black}{[#2]}}}
\newcommand{\remove}[1]{\ifbool{togglechanges}
    {}    
    {\textcolor{red!70!black}{#1}}}
\newcommand{\inset}[1]{\ifbool{togglechanges}
    {#1}  
    {\textcolor{green!50!black}{#1}}}
\newcommand{\optional}[1]{\ifbool{togglechanges}
    {#1}  
    {\textcolor{yellow!50!orange!80!gray}{#1}}}
\newcommand{\citeremind}[1]{%
    [\textcolor{blue!75!black!80!yellow}{
        $\blacksquare$%
           \ifthenelse{\isempty{#1}}
               {}
               {\textsuperscript{\textsf{#1}}}%
        }]\xspace}
\newcommand{\todo}[1]{
    \textcolor{orange!80!yellow!95!black}{\textbf{[}%
        \ifthenelse{\isempty{#1}}%
        {\text{$\blacksquare$}}%
        {{\small\textsf{#1}}}%
        \textbf{]}}}
        
\newcommand{\mitaffil}{\footnotesize Research Laboratory of Electronics and Department of Physics, Massachusetts Institute of Technology, Cambridge, Massachusetts 02139, USA}
\newcommand{\harvardaffil}{\footnotesize Harvard John A. Paulson School of Engineering and Applied Sciences, Harvard University, Cambridge Massachusetts 02138, USA}
\newcommand{\isnaffil}{\footnotesize Institute for Soldier Nanotechnologies, Massachusetts Institute of Technology, Cambridge, Massachusetts 02139, USA}
\newcommand{\technionaffil}{\footnotesize Department of Electrical and Computer Engineering, Technion–Israel Institute of Technology, 32000 Haifa, Israel}

\myexternaldocument{SM}

\begin{document}

\title{Observation of enhanced free-electron radiation from photonic flatband resonances}

\author{Yi~Yang}
\email{yiy@mit.edu}\email{chrc@mit.edu}
\thanks{Y.~Y. and C.~R.-C. contributed equally to this work.}
\affiliation{\mitaffil}
\author{Charles~Roques-Carmes}
\email{yiy@mit.edu}\email{chrc@mit.edu}
\thanks{Y.~Y. and C.~R.-C. contributed equally to this work.}
\affiliation{\mitaffil}
\author{Steven~E.~Kooi}
\affiliation{\isnaffil}
\author{Haoning~Tang}
\affiliation{\harvardaffil}
\author{Justin~Beroz}
\affiliation{\mitaffil}
\author{Eric~Mazur}
\affiliation{\harvardaffil}
\author{Ido~Kaminer}
\affiliation{\technionaffil}

\author{John~D.~Joannopoulos}
\affiliation{\mitaffil}
\affiliation{\isnaffil}
\author{Marin~Solja\v{c}i\'{c}}
\affiliation{\mitaffil}
\affiliation{\isnaffil}

\date{\today} 

\begin{abstract}

\end{abstract}

\maketitle


\textbf{
Flatbands emerge from a myriad of structures such as Landau levels~\cite{landau1930Diamagnetism}, Lieb and Kagome lattices~\cite{mukherjee2015observation,vicencio2015observation,slot2017experimental,kang2020dirac}, line graphs~\cite{kollar2019line}, and more recently moir\'{e} superlattices~\cite{cao2018unconventional,wang2020localization}. They enable unique properties including slow light~\cite{baba2008slow} in photonics, correlated phases~\cite{cao2018unconventional} in electronics, and supercollimation~\cite{rakich2006achieving,park2008electron} in both systems.
Despite these intense parallel efforts, flatbands have never been shown to affect the core light-matter interaction between electrons and photons, which is limited by a dimensionality mismatch.
Here, 
we reveal that a photonic flatband can overcome this mismatch between localized electrons and extended photons and thus remarkably boost their light-matter interaction.
We design flatband resonances in a silicon-on-insulator photonic crystal slab to control and enhance the radiation emission from free electrons by tuning their trajectory and velocity.
In particular, we record a 100-fold radiation enhancement from the conventional diffraction-enabled Smith--Purcell radiation~\cite{smith1953visible}, and show the potential of our approach to achieve $10^6$-fold enhancements and beyond.
The enhancement also enables us to perform polarization shaping of free electron radiation from multiple flatbands and demonstrate an approach to measure photonic bands via angle-resolved electron-beam measurements.
Our results suggest flatbands as ideal test beds for strong light-electron interaction in various systems, with particular relevance for efficient and compact free-electron light sources and accelerators.
}

The interaction between free electrons and optical environments, foundational in modern electron microscopy and spectroscopy~\cite{Polman2019,de2010optical}, gives rise to a multitude of radiative processes~\cite{Polman2019,de2010optical,friedman1988spontaneous,schachter2011beam,cherenkov34,smith1953visible,kimura1995laser,mizuno1987experimental}.
These processes constitute an invaluable diagnostic platform, however, usually at low coupling strength, because of the limited interaction cross-section.
Accordingly, longer interaction lengths in extended structures~\cite{luo2003cerenkov,de2003cherenkov,lin2018controlling,adamo2009light,kaminer2017spectrally} can achieve stronger coupling strength, 
as exemplified by a thousand-photon stimulated emission and absorption by a single electron~\cite{dahan2020resonant}, and thousand-electron-volt acceleration shown in an integrated dielectric laser accelerator~\cite{sapra2020chip}.
Free electrons propagating in extended periodic structures can emit Smith-Purcell radiation, an effect resulting from the diffraction of the electrons' near-field. It has been proposed that coupling free electrons with nanophotonic Bloch resonances can result in exotic radiation regimes~\cite{luo2003cerenkov} and strong radiation enhancements \cite{yang2018maximal, yamaguti2002photonic} with the prospect of being integrated in optimized nanophotonic structures \cite{roques2019towards,liu2017integrated,haeusler2021boosting}. Such enhancements may result in regimes of free-electron stimulated emission and lasing \cite{Rivera:20, pellegrini2016physics, urata1998superradiant,andrews2004gain,kumar2006analysis, freund2004linearized, schachter1989smith}.

The key rule depicting what radiation is emitted from free electrons and at what efficiency is phase matching~\cite{friedman1988spontaneous,schachter2011beam}: it requires that the electron velocity $\mathbf{v}$ and the photon phase velocity $\omega/\mathbf{k}$ together satisfy $\omega=\mathbf{v}\cdot\mathbf{k}$. Importantly, the phase-matching condition only involves the longitudinal momentum of the emitted radiation. There remains a continuum of transverse momenta simultaneously allowed in the emitted field. 
However, photonic modes are not necessarily supported at all of these momenta, which results in a transverse momentum mismatch.
Many previous theoretical works circumvent this mismatch by dimension reduction, \eg assuming electron sheets~\cite{sheetbeam} as a mathematical convenience that reduces computational complexity. In this case, the transverse momentum is neglected and the mismatch highlighted above disappears. Because of this theoretical simplification, the predicted radiation falls short in realistic three-dimensional settings.

Instead, to ultimately achieve a full energy–momentum matching with point electrons, we propose to introduce a continuum of photonic modes along the transverse momenta of free electrons, which turns out to be the signature feature of flatbands.
There are various ways to create photonic flatbands~\cite{leykam2018artificial,leykam2018perspective}, such as laser-written waveguide arrays~\cite{mukherjee2015observation,vicencio2015observation,tang2020photonic}, photonic crystals (PhCs)~\cite{li2008systematic,rakich2006achieving,baba2008slow}, moir\'{e} structures~\cite{wang2020localization,lou2021theory,dong2021flat,nguyen2021magic,tang2021modeling}, and their non-Hermitian counterparts~\cite{leykam2017flat,pan2018photonic}.
Their applications include lasers~\cite{noda2017photonic,longhi2019photonic}, imaging~\cite{xia2018unconventional}, and optical communications~\cite{baba2008slow}.
Despite these advances, photonic flatbands have been mostly studied under pure optical settings, while never explored for electron-photon interactions. 

In this work, we realize a full energy–momentum matching between free electrons and a continuum of photonic flatband resonances supported by a silicon-on-insulator PhC slab.
The phase matching is realized by tuning the velocity of free electrons and their twist angles relative to the nearby PhC slab.
By leveraging the flatband scheme, we experimentally observe 100-fold enhancement of Smith--Purcell radiation.
We provide three confirmations for the flatband origin of the radiation---band structure measurements with laser beams and electron beams, respectively, as well as polarization shaping of free-electron radiation from multiple photonic flatbands.

\begin{figure}[htbp]
 \includegraphics[width=1\linewidth]{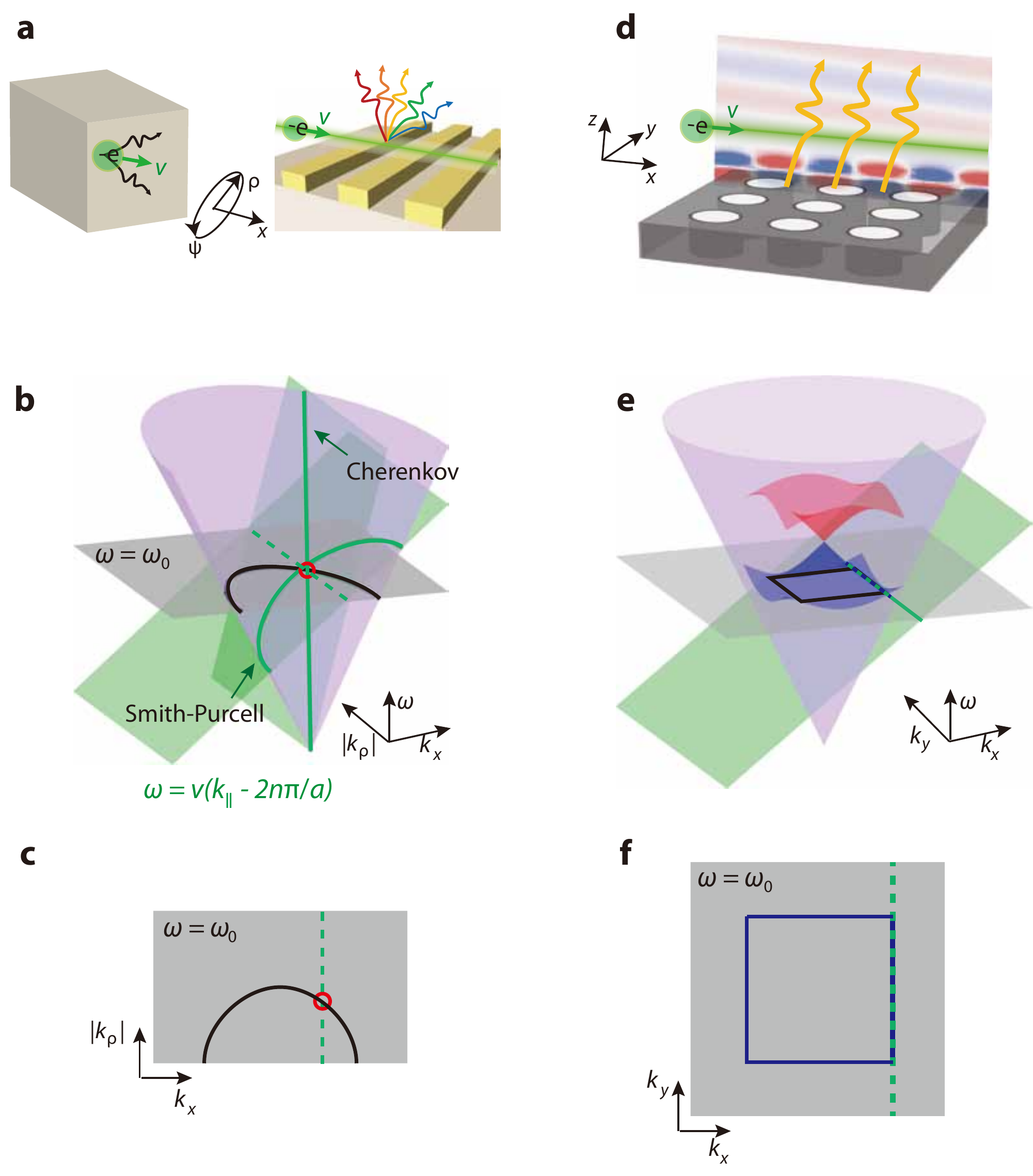}
        \caption{
        \textbf{Flatband-resonance-mediated free-electron radiation.}
        \textbf{a.} Conventional effects of free-electron radiation. Free electrons in a bulk medium emit Cherenkov radiation when satisfying the superluminal condition (left) or emit Smith--Purcell radiation via diffraction from periodic structures (right).
        \textbf{b.} In momentum space, Cherenkov and Smith--Purcell radiation occur at frequencies and momenta where the light cone (purple; $\omega=ck/n$) and the electron surface (green; $\omega=v(k_x-2n\pi/a$) intersect. 
        \textbf{c.} At a specific frequency $\omega=\omega_0$, their intersection is always a point degeneracy (red circle), which corresponds to the Cherenkov cone or the Smith--Purcell dispersion relation.
        \textbf{d.} By leveraging resonances in PhC slabs, free-electron radiation can be substantially modified and enhanced.
        \textbf{e.} Because the transverse momentum (green line on green surface) of the electron surface is unbounded, a continuum of flat resonances (blue square) is crucial for maximizing free-electron-light interaction.   
        \textbf{f.} At a certain velocity and frequency, a line degeneracy exactly at the flatband, can be formed between the photonic bands and the electron surface.
        }
\label{fig:concept}
\end{figure}

\begin{figure*}[htbp]
 \includegraphics[width=\linewidth]{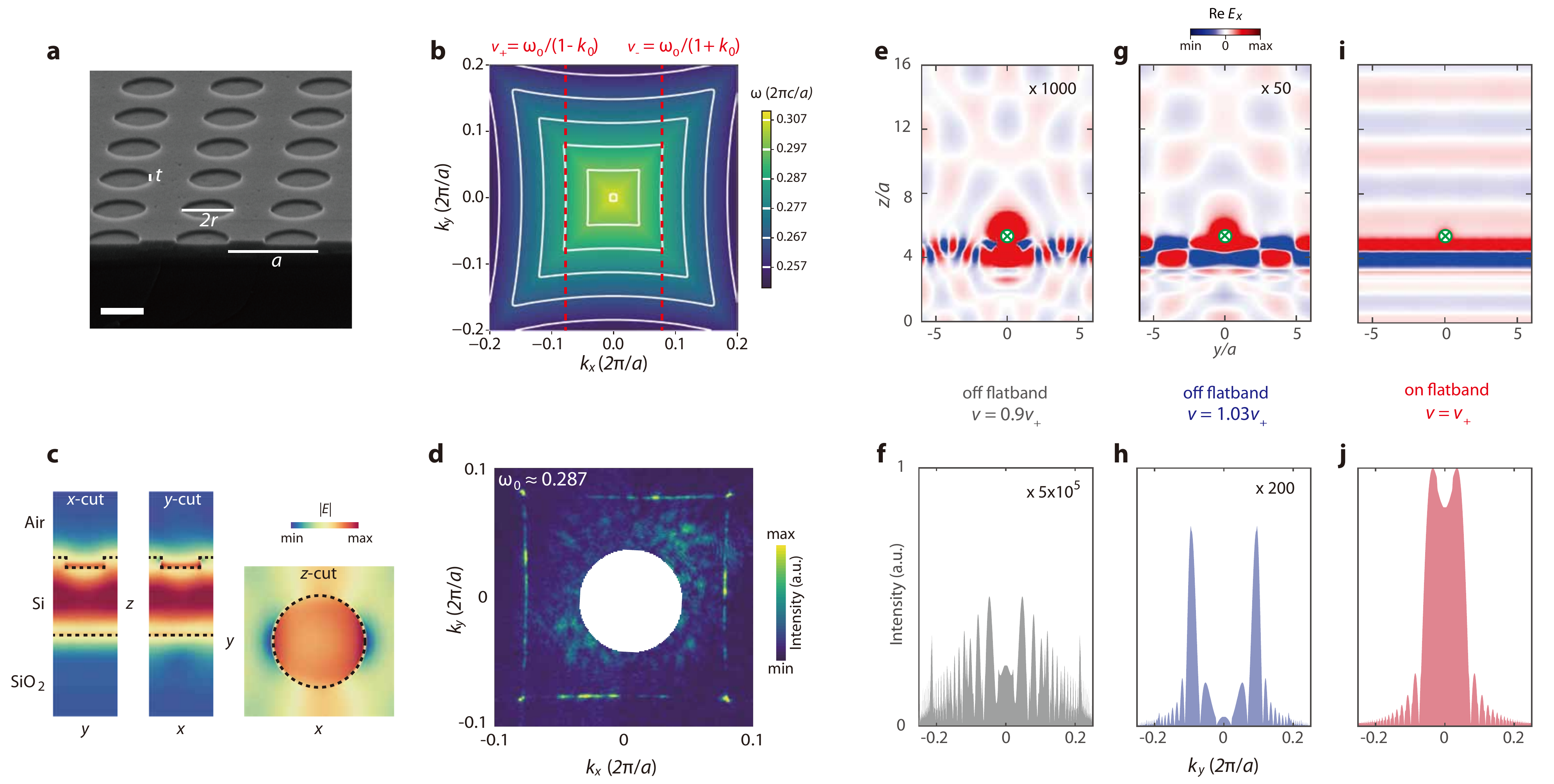}
        \caption{
        \textbf{Photonic flatbands for free electron radiation.}
        \textbf{a.} Scanning electron micrograph of the PhC slab. Scale bar \SI{200}{\nm}.
        \textbf{b.} Predicted isofrequency contours. 
        A flatband appears near $\omega_0\approx0.287$ and $\abs{k_{x}}=k_0\approx0.08$. Under resonant electron velocities $v_\pm$ (dashed red lines; Eq.~\ref{eq:beta}), maximal overlap appears between the electron surface and the photonic band. 
        \textbf{c.} Mode profiles of $|E|$ at $(k_x,k_y)=(k_0,0)$, the center of the flatband. $x-$ and $y-$cuts are shown at the center of the unit cell. $z-$cut is chosen in the middle of the air hole. Dashed lines indicate material boundaries.
        \textbf{d.} Measured isofrequency contours at the flatband (see Sec.~S3 for details).
        \textbf{e-j.} Simulated radiation field profiles $\Re{E_x}$ (top) and the associated Fourier analysis (bottom) at $\omega_0$ when the electron surface is off (e and f: $v=0.9v_+$; g and h: $v=1.03v_+$;) and on (i and j; $v=v_+$) the flatband at $\omega_0$. Enhanced plane-wave-like radiation can be achieved with the flatband. The green circles indicate the location of the point free electrons.
        In e and f, radiation is dominated by diffraction, as indicated by the Fourier components towards all transverse directions.
        In g and h, radiation is mainly contributed from discrete resonances at $\abs{k_y}\approx0.1$ (also see Fig.~S1c and d).
        In i and j, the flatband strongly excites a continuum of Fourier components $k_y$ of the far-field radiation.
        Sidelobes appear because of the finite-size effects of the simulations.
        }
\label{fig:bands}
\end{figure*}


The transverse mismatch puts a limit on all conventional free-electron radiation mechanisms. Consider the quintessential Cherenkov~\cite{cherenkov34} and Smith--Purcell~\cite{smith1953visible} effects. Free charged particles, such as electrons, generate Cherenkov radiation if their velocity $v$ exceeds the phase velocity of light $c/\sqrt{\epsilon}$ in a homogeneous medium with permittivity $\epsilon$ (Fig.~\ref{fig:concept}a left). 
They can also generate Smith--Purcell radiation near periodic structures via diffraction of their near field (Fig.~\ref{fig:concept}a right).
Assuming an electron propagates in the $x$ direction, \ie $\mathbf{v}=v \hat{\mathbf{x}}$, its external fields obey $\omega=v(k_x - 2n\pi/a)$ (referred to as the electron surface hereafter; green surfaces in Fig.~\ref{fig:concept}b), where $a$ is periodicity, $n=0$ for Cherenkov radiation, and $n\in\mathbb{Z}_{\neq0}$ is the diffraction order for Smith--Purcell radiation.
In cylindrical coordinates $(x,\rho,\psi)$ where $k_x$ and $k_\rho\hat{\rho}=k_y\hat{y}+k_z\hat{z}$ are momenta parallel and perpendicular to the electron velocity, respectively, the intersections between the electron surfaces and the light cone $\omega = ck/\sqrt{\epsilon}$ (purple in Fig.~\ref{fig:concept}b) dictate the dispersion relations of both types of radiation.
Specifically, Cherenkov radiation is spectrally continuous while Smith--Purcell radiation forms discrete spectral windows under various diffraction orders (Fig.~\ref{fig:concept}b).
However, in both cases, emission at each frequency $\omega_0$ can only occur at discrete momenta, \ie the phase-matching condition is only satisfied at point degeneracies (red circles in Fig.~\ref{fig:concept}c) between the electron surface and the isofrequency contour.

Such nonresonant bare point degeneracies are not optimal for maximizing electron-light interaction. 
To go beyond that, first, one can introduce photonic resonances, instead of bare diffraction to free-space plane waves, and couple them with free electrons to obtain enhancement in tandem with external coupling~\cite{yang2018maximal}. 
Second, one can create higher degeneracies in momentum space, \ie line degeneracies, between the electron surfaces and a continuum of photonic resonances to further enhance their interaction.
Such conditions can be met by considering the interaction between free electrons and photonic flatbands.
In fact, it was theoretically predicted that radiated power enhancements could occur near discrete momenta where the transverse component (orthogonal to the electron trajectory) of group velocity vanishes~\cite{kremers2009theory}.
Rather than discrete momenta, flatbands naturally satisfy this condition over a wide momentum bandwidth that should further improve the enhancement.

Specifically, we consider Smith--Purcell radiation from a PhC slab that hosts flatband resonances, as shown in Fig.~\ref{fig:concept}d. For conceptual clarity, we first ignore its non-Hermiticity and discuss the band structure of its Hermitian counterpart that is continuously translationally invariant along the $z$ direction. Near the center of the Brillouin zone ($\Gamma$ point), the PhC supports bands that can be described by an effective two-dimensional, square-lattice, Dirac Hamiltonian~\cite{chiu2014classification} 
\begin{align}
    h(k_x,k_y) = \omega_{\rm{d}}\sigma_0+\nu\sin k_+\sigmax+\nu\sin k_-\sigmay + m\sigmaz,
    \label{eq:ham}
\end{align}
where $\omega_{\rm{d}}$ is the frequency of degeneracy, $k_\pm=(k_x \pm k_y)/\sqrt{2}$, $\sigma_{0}$ is the identity matrix, and $\sigma_{x,y,z}$ are Pauli matrices. Its associated bands are shown by the red and blue surfaces above the light cone in Fig.~\ref{fig:concept}e.
For negligible mass $m\approx 0$, this Hamiltonian hosts a flatband at $\left(\omega_0,\mathbf{k}_0\right)$ (Fig.~\ref{fig:concept}e and f), where $\omega_0 \equiv \omega_{\rm{d}}\pm \nu$ and the isofrquency contour $\mathbf{k}_0$ becomes flat, at the critical point between elliptic and hyperbolic dispersions. 
The frequency of the flatband $\omega_0$ and the size of its contour $\mathbf{k}_0$ can both be tuned by proper design.
We need to control simultaneously the velocity of electrons and their in-plane, twist angle relative to the PhC (see Sec.~S1) to realize the desired single-frequency line degeneracy (dashed green-blue line in Fig.~\ref{fig:concept}f) between the electron surface and the flatband. 
For the flatband perpendicular to the $\Gamma-X$ direction (Fig.~\ref{fig:concept}f), the optimal twist angle is zero degree. 
At other non-optimal velocities and twist angles, only point degeneracies can be formed at a single frequency (see~Sec.~S1). 

\begin{figure*}[htbp]
 \includegraphics[width=0.9\linewidth]{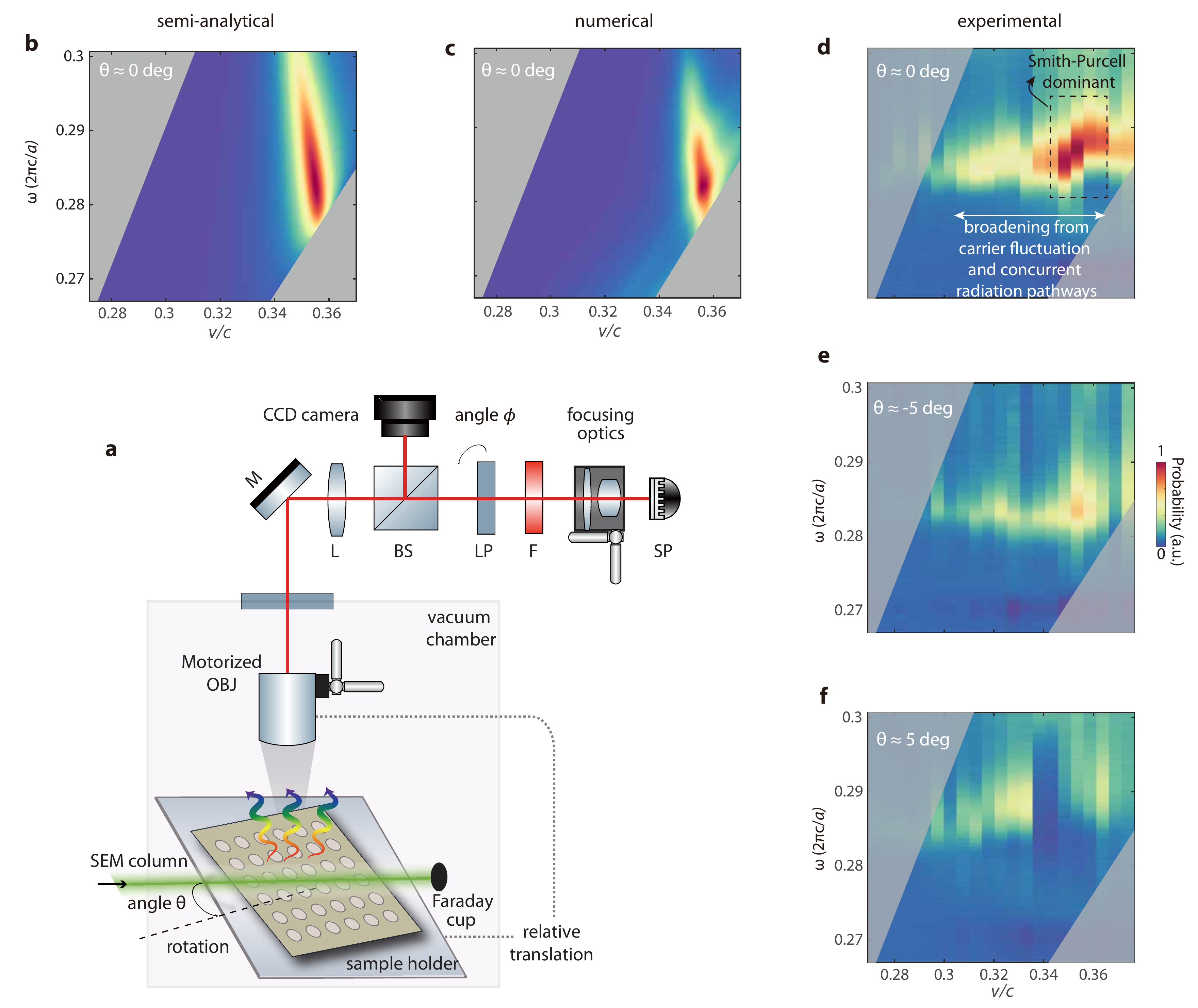}
        \caption{
        \textbf{Measurement of radiation from flatbands.}
        \textbf{a.} Modified scanning-electron-microscope experimental setup, used to analyze free-electron radiation from the PhC slab with a camera and a spectrometer (for details see~Sec.~S2).
        Radiation was measured under three in-plane twist angles $\theta$ between the electron beam and the PhC slab.
        The linear polarizer was oriented at angle $\phi\approx\SI{0}{\degree}$, \ie parallel to the electron beam, consistent to the far-field polarization of the mode.
        A low-pass frequency filter with $\omega\lesssim 0.31$ (cut-off wavelength $\SI{1400}{\nm}$) was used to reduce incoherent cathodoluminescence near silicon's electronic band gap.
        OBJ, objective; M, mirror; L, lens; BS, beam splitter; LP, linear polarizer; F, filter; SP, spectrometer.
        \textbf{b-c.} Predicted radiation probability from semi-analytical (b; Sec.~S5) and numerical (c; Sec.~S6) calculations for point electrons perfectly parallel to the sample surface. Maximal enhancement appears near the flat-band resonances. Shaded area, in both theoretical and experimental results, indicate parameter space outside the numerical aperture of the objective. 
        \textbf{d-e.} Measured radiation probability at $\theta\approx\SI{0}{\degree}$ (d), $\theta\approx\SI{-5}{\degree}$ (e), and $\theta\approx\SI{5}{\degree}$ (f) in-plane twist angles.
        }
\label{fig:rad}
\end{figure*}

To realize such flatbands experimentally, we designed and fabricated an SOI, square-lattice, PhC slab (Fig.~\ref{fig:bands}a). The macroscopic-scale PhC slab, fabricated with interference lithography, has a periodicity of $a\approx\SI{430}{\nm}$. The thicknesses of the device layer and the buried oxide layer are $1.19a$ and $2.38a$, respectively. The air holes of the PhC slab have a radius of $r\approx0.34a$ and a depth of $t\approx0.13a$ (Fig.~\ref{fig:bands}a). We choose to etch the air holes in this shallow manner so that non-Hermiticity appears perturbatively, without substantially modifying the desired dispersion of the real part of the complex frequencies [Fig.~\ref{fig:concept}e and f]. Therefore, the effective Hamiltonian in Eq.~\eqref{eq:ham} described above still applies. 

We first focus on the lowest, TM-like band above the light cone whose isofrequency contour is shown in Fig.~\ref{fig:bands}b. 
A flatband around $\omega_0\approx0.287$ (in units of $2\pi c/a$) appears near $\abs{k_{x}}=k_0\approx0.08$ (in units of $2\pi/a$) along the $k_y$ direction and its \SI{90}{\degree}-rotation partners. 
In the second Brillouin zone, the overlap between the electron surface and the flatband can appear at $k_x=\mp k_0$, respectively, under two discrete velocities $v_\pm$ in the $x$ direction (Fig.~\ref{fig:bands}b)
\begin{align}
    v_{\pm} = \omega_0/\left(1\mp k_0\right).
    \label{eq:beta}
\end{align}
The electric mode profiles $|E|$ within a unit cell at $\omega_0$ and $(k_x,k_y)=(k_0,0)$ are shown in Fig.~\ref{fig:bands}c. 
Since the mode of interest originates from a perturbed guided mode in the slab, it remains mostly confined within the device Si layer, while the shallow air holes provide external radiative coupling.
The existence of the flatband near $\omega_0$ was confirmed with Fourier scattering spectroscopy measurements (see~Sec.~S3).
Its quality factor $Q_{\rm{o}}$ from this optical characterization was measured as $Q_{\rm{o}}\approx400$, as shown in Fig.~S3.

The flatband modifies and enhances free-electron radiation substantially, as shown by the comparison between the simulated radiation patterns and the associated Fourier analysis shown in Fig.~\ref{fig:bands}e to j (see numerical methods in~Sec.~S6).
We identify three regimes of radiation generation.
First, at frequencies where the electron surface is far away from the photonic band (Fig.~\ref{fig:bands}e and f), resonant effects are negligible. Still, radiation is generated by diffraction and allowed towards all transverse directions (Fig.~\ref{fig:bands}f), similar to the conventional Smith--Purcell radiation (Fig.~\ref{fig:concept}b).
Second, at frequencies near discrete resonances, \ie point degeneracies between the free electron surface and the photonic band, radiation toward discrete directions gets selectively enhanced~\cite{yamaguti2002photonic} (see Fig.~S1c-d).
This is confirmed by the associated, discrete Fourier peaks near $\abs{k_y}\approx0.1$ (Fig.~\ref{fig:bands}h), originating from the tail of a resonance at a lower frequency.
Finally, when the electron surface intersects (Fig.~\ref{fig:bands}i and j) the flatband (phase matching condition shown in Fig.~\ref{fig:concept}f and Fig.~\ref{fig:bands}b), enhanced, plane-wave-like emission is obtained despite the point nature of the electron source.
The associated far-field Fourier transform reveals the excitation of the flatband continuum $\abs{k_y}\lesssim0.08$.
Taken together, compared to those when the electron surface is off the flatband by 10\% (3\%), the flatband is predicted to achieve a 1000-fold (50-fold) field enhancement, \ie a $10^6$ ($10^3$) intensity enhancement.

We measured such predicted flatband-enhanced, free-electron radiation (Fig.~\ref{fig:rad}).
In our setup (Fig.~\ref{fig:rad}a and see~Sec.~S2), free electrons passed above the PhC at a grazing angle ($\approx\SI{1}{\degree}$) under \SIrange{20}{40}{\kV} acceleration voltages from a LaB$_6$ electronic gun in the high-current regime $\approx\SIrange{30}{60}{\micro\ampere}$.
We chose electron velocities near $v_+$, rather than $v_-$ because the electron gun produces higher-quality beams under higher voltages~(see~Sec.~S4).
We tuned the velocity of free electrons to form different intersections between the electron surface and the photonic band and recorded the associated radiation.
The control of the in-plane, relative orientation, \ie a twist angle $\theta$ (Fig.~\ref{fig:rad}a), between the PhC slab and the electron beam enabled us to create radiation probability maps $P(v,\omega;\theta)$ as a function of electron velocity $v$ and frequency $\omega$. 
Fig.~\ref{fig:rad}b and c show the predicted radiation based on semi-analytical (see~Sec.~S5) and numerical (see Sec.~S6) calculations at a zero in-plane angle $\theta$. 
Both of the theoretical results indicate a region of radiation enhancement near $(v,\omega)\approx(0.35c,0.285)$ where the electron surface intersects the flatband resonances.

These predictions were confirmed in our $\theta=\SI{0}{\degree}$ data (see measurement protocol in Sec.~S2D) in Fig.~\ref{fig:rad}d. 
We extracted a quality factor $Q_{\rm{e}}\approx50$ from the emission peak and its linewidth in these electron radiation measurements. $Q_{\rm{e}}$ was substantially higher than the quality factor of the conventional Smith--Purcell radiation $Q_{\rm{SP}}\equiv c/v\approx3$, and thus indicates the resonant enhancement.
Also in Fig.~\ref{fig:rad}d, we quantified a $10^2$-fold enhancement (see Sec.~S7), which is defined as the ratio between the peak radiation and the average radiation in the region $(v<0.32c,\omega<0.28)$, about $10\%$ off the flatband as the case shown in Fig.~\ref{fig:bands}e and f.

We observed certain deviations from the theory predictions, specifically, shift and broadening.
The measured $Q_{\rm{e}}$ was lower than $Q_{\rm{o}}$ from the optical measurements (see Sec.~S3), and the measured enhancement was lower than the prediction in Fig.~\ref{fig:bands}.
The reduction could stem from a few experimental uncertainties (see Sec.~S8), chief among which is the increase of free carriers in silicon under electron-beam exposure (see Sec.~S8).
To account for the effective doping of carriers, the refractive index of the top silicon layer is chosen as $n_{\rm{Si}}=3.2+0.03\iu$ (corresponding to a doping concentration \SIrange{e20}{e21}{\per\cm\cubed}; see Sec.~S8) in both Fig.~\ref{fig:rad}b and c.
Notably, the measured enhancement region is broadened (Fig.~\ref{fig:rad}d annotations) along the velocity axis compared to that in the theoretical calculations (Fig.~\ref{fig:rad}b and c). 
Such broadening could be caused by two major reasons. 
First, the temporal variation and spatially inhomogeneity of carriers in silicon (see Sec.~S8) render the adoption of a constant refractive index $n_{\rm{Si}}$ only a zeroth-order approximation.
Second, in contrast to the theoretical considerations (Fig.~\ref{fig:rad}b and c) of Smith-Purcell radiation where electrons are assumed perfectly parallel to the sample surface, in the experiment the grazing-angle electron beam inevitably impinged onto the PhC slab, causing extra radiation pathways~\cite{brenny2014quantifying}, including incoherent cathodoluminescence and transition radiation, which are less sensitive to electron velocity and can be simultaneously enhanced by the flatband resonance. 
The associated supporting evidence is the fact that the observed additional emission outside the Smith--Purcell dominant region still appears near the flatband frequency (see annotations in Fig.~\ref{fig:rad}d).

\begin{figure}[htbp]
 \includegraphics[width=1\linewidth]{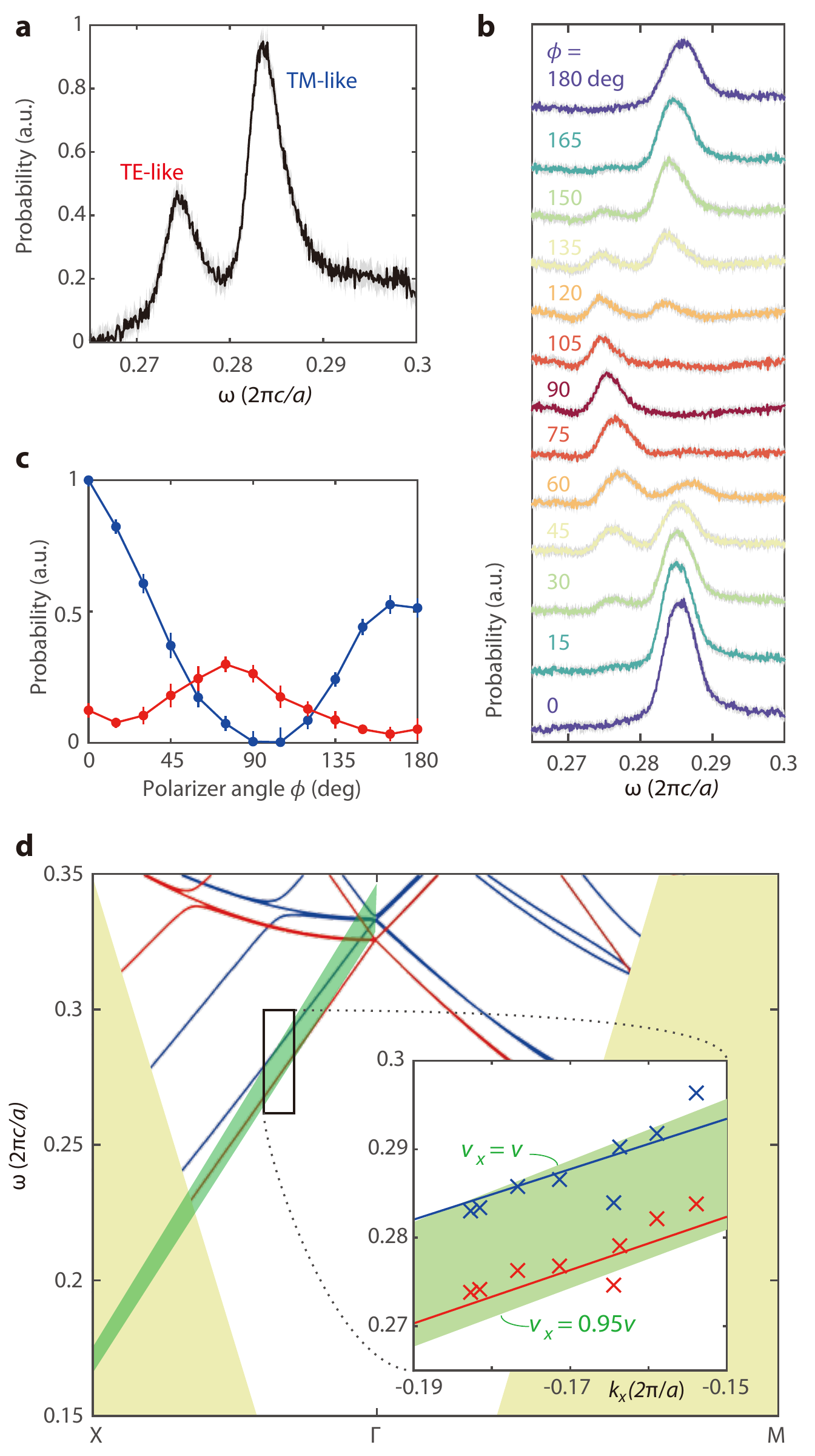}
        \caption{
        \textbf{Polarization control of free-electron radiation.}
        \textbf{a-b.} Simultaneous excitation of multiple flatbands (a: without the detection polarizer) and their contrasting polarization dependence (b: with polarizer, pass angle $\phi\in\left[\SI{0}{\degree},\SI{180}{\degree}\right]$).    
        \textbf{c.} Peak radiation probability of the TM-like (blue) and TE-like (red) modes under various $\phi$.
        \textbf{d.} Photonic bands  (theory: lines) measured (crosses) with free-electron radiation by translating the field of view of the collection optics.
        In this set of measurements, the electron velocity is $v\approx0.343c$.
        In a-c, shadings and error bars indicate standard error $\pm1\sigma$. Spectra are offset for readability in b.         
        In d, shaded green indicates the longitudinal cut of the electron surface $\omega=v k_x$ under 5\% in-plane velocity uncertainty. Shaded yellow is outside the light cone.
        }
\label{fig:pol}
\end{figure}

We next explore radiation as a function of the twist angle $\theta$. 
Since the flatband is one-dimensional, \ie only flat along the $k_y$ direction, the associated radiation enhancement should depend crucially on even a small twist of the in-plane angle $\theta$ while regular Smith--Purcell radiation does not.
Such contrasting dependence was also confirmed by the twist angle $\theta\approx\pm\SI{5}{\degree}$ measurements in Fig.~\ref{fig:rad}e and f, where the enhancement reduces to about 30-fold, lower than that in Fig.~\ref{fig:rad}d.

Finally, we reveal how resonance-mediated free-electron radiation can be utilized to control the polarization of free-electron emission. 
The conventional Smith-Purcell radiation has a preferred polarization parallel to the electron beam, which calls for the ability to generate and control arbitrary polarization states. Towards this direction, recent endeavors are epitomized by bianisotropic metasurfaces~\cite{wang2016manipulating, jing2021polarization}, albeit limited to the microwave regime.
Since our PhC slab was shallowly etched, the frequency splitting between TE- and TM-like bands is moderate, which enables simultaneous excitation of the two bands using free electrons at certain velocities. This is confirmed by the measured two-peak spectrum at $v\approx0.343c$ (Fig.~\ref{fig:pol}a).
The two peaks correspond to different $s$- and $p$- far-field polarization---the TE- and TM-like peaks exhibited maximum emission at a polarization orthogonal and parallel to the electron beam direction, respectively~\cite{signalReduction}.

Furthermore, the simultaneous excitation of the two modes enables us to measure their dispersion from their far-field radiation. Since the electron surface is almost parallel to the two bands under the chosen velocity (\cf~bands and the electron surfaces in Fig.~\ref{fig:pol}d), the overlapped photonic band structure can be extracted from the emission spectra by translating the objective's field of view (Fig.~\ref{fig:rad}a and see~Sec.~S5 for details). %
In the experimentally accessible region (inset in Fig.~\ref{fig:pol}d) of the band structure, we observed a linear, monotonic trend of dispersion of both modes (crosses), in accordance with the calculated bands (lines).

To sum up, we have demonstrated a full energy-momentum matching between free electrons and an optical environment via a continuum of flatband resonances.
The flatband gives rise to strongly enhanced free-electron radiation, which also enables us to perform polarization shaping and measure photonic band structures via free electrons.
The enhancement from flatband can be further improved by theoretical and experimental means. 
Beyond this work where free electrons couple to one-dimensional flatband resonances, higher-dimensional flatbands~\cite{tang2021modeling} and bound states in the continuum~\cite{cerjan2019bound,cerjan2021observation} could be designed to increase the enhancement and its robustness against the angular spread of electron beams and fabrication disorder.
Experimentally, ultrafast scanning transmission electron microscopes~\cite{wang2020coherent,sapra2020chip} and integrated free-electron emitters~\cite{guerrera2016nanofabrication,liu2017integrated}
could provide more collimated and focused electron beams to reduce material deterioration, increase interaction duration, and pinpoint the flatband coupling condition more accurately.

The flatband-mediated electron-light interaction, realized here with PhCs, can be generically applied to a variety of material platforms (\eg two-dimensional, plasmonic, and hybrid materials) and spectral ranges (\eg THz, ultraviolet, and X-ray generation) in both the classical and quantum regimes.
In particular, the flatband approach could be useful for building more efficient integrated radiation sources~\cite{liu2017integrated} and particle accelerators~\cite{sapra2020chip,zhao2020design,shiloh2021electron}.
More broadly, our results highlight free electrons as pumps~\cite{fallah2021nonreciprocal} and probes~\cite{peng2019probing,yu2019transition} for photonic topological bands and defects. By varying the velocity of free electrons and their twist angles relative to photonic structures, coupling with boundary modes can be realized and closed contours constructed in momentum space to diagnose bulk properties such as winding numbers.

\section{Acknowledgements}
The authors thank Tim Savas for fabricating the samples; Aviram Massuda and Jamison Sloan for contributions to the setup building; Di Zhu and Marko Lon\v{c}ar for equipment sharing; and Chenkai Mao, Owen D. Miller, and Nicholas Rivera for stimulating conversations.
This material is based upon work supported in part by the U.S. Army Research Office through the Institute for Soldier Nanotechnologies under contract number~W911NF-18–2–0048, the Air Force Office of Scientific Research under the award number FA9550-20-1-0115 and FA9550-21-1-0299, and the U.S. Office of Naval Research (ONR) Multidisciplinary University Research Initiative (MURI) Grant No. N00014-20-1-2325 on Robust Photonic Materials with High-Order Topological Protection.
C.~R.-C. acknowledges funding from the MathWorks Engineering Fellowship Fund by MathWorks Inc.

\section{Author contributions}
Y.~Y., C.~R.-C., I.~K., and M.~S. conceived the project.
Y.~Y. designed the sample.
C.~R.-C. and S.~E.~K. performed the radiation measurements.
H.~T. and Y.~Y. performed the Fourier scattering spectroscopy. 
J.~B. designed and fabricated the objective motorized stage with inputs from C.~R.-C. and S.~K..
Y.~Y. and C.~R.-C. analyzed the data.
Y.~Y. and C.~R.-C. wrote the manuscript with inputs from all authors.
E.~M., I.~K., J.~D.~J., and M.~S. supervised the project.

\providecommand{\noopsort}[1]{}\providecommand{\singleletter}[1]{#1}%


\begin{thebibliography}{69}
\bibitem [{sig()}]{signalReduction}%
  \BibitemOpen
  \href@noop {} {}\bibinfo {note} {The signal reduction at
  $\phi=\SI{180}{\degree}$, compared to that at $\phi=\SI{0}{\degree}$, could
  result from the charge accumulation on the sample under longer beam exposure,
  since we measured $\phi=\SIrange{0}{180}{\degree}$ sequentially with
  increasing angle and with a high-current excitation.}\BibitemShut {Stop}%
\end{thebibliography}
\end{document}